\documentclass[conference]{IEEEtran}

\usepackage[dvips]{graphicx}
\graphicspath{{./figures/}}
\DeclareGraphicsExtensions{.eps}
\usepackage{float}

\usepackage[cmex10]{amsmath}
\usepackage{amsfonts}
\usepackage{mathtools}
\usepackage{amssymb}

\usepackage{algorithmic}						
\usepackage{algorithm}
\algsetup{linenosize=\scriptsize}
\makeatletter
\newcommand\fs@norules{\def\@fs@cfont{\bfseries}\let\@fs@capt\floatc@ruled
  \def\@fs@pre{}%
  \def\@fs@post{}%
  \def\@fs@mid{\kern3pt}%
  \let\@fs@iftopcapt\iftrue}
\makeatother
\floatstyle{norules}
\restylefloat{algorithm}

\usepackage[tight,footnotesize]{subfigure}
\usepackage[font=small,skip=5pt]{caption}
\usepackage{wrapfig}
\usepackage{lipsum}

\usepackage[usenames, dvipsnames]{color}

\usepackage{times} 

\renewcommand{\theenumii}{\theenumi.\arabic{enumii}.}
\usepackage{enumitem}
\setitemize{noitemsep,topsep=0pt,parsep=0pt,partopsep=0pt}
\usepackage{tikz}
\usepackage{pgf-umlsd}
\usepgflibrary{arrows} 
\usepackage{listings}

\newcommand{\system}{$\operatorname{FogMQ}$}
\newcommand{\redis}{$\operatorname{Redis}$}
\newcommand{\nats}{$\operatorname{Nats}$}
\newcommand{\zmq}{$\operatorname{ZeroMQ}$}
\newcommand{\cgroups}{$\operatorname{cgroups}$}
\newcommand{\protocol}{$\operatorname{Flock}$}
\newcommand{\linpack}{$\operatorname{Linpack}$}

\newtheorem{theorem}{Theorem}[section]
\newtheorem{lemma}[theorem]{Lemma}

\newcommand{\qed}{\nobreak \ifvmode \relax \else
      \ifdim\lastskip<1.5em \hskip-\lastskip
      \hskip1.5em plus0em minus0.5em \fi \nobreak
      \vrule height0.75em width0.5em depth0.25em\fi}

\newcommand{\comment}[1]{ }

\usepackage[printonlyused]{acronym}
\acrodef{vm}[VM]{Virtual Machine}
\acrodef{iot}[IoT]{Internet of Things}
\acrodef{rtt}[RTT]{Round Trip Time}
\acrodef{ec}[EC]{Edge Cloud}
\acrodef{ne}[NE]{Nash Equilibrium}
\acrodef{mdp}[MDP]{Markov Decision Process}
\acrodef{poa}[PoA]{Price of Anarchy}
\acrodef{cdf}[CDF]{Cumulative Distribution Function}

\newcommand{\Rset}{\mathbb{R}} 

\newcommand{\migrate}[3]{#2 \xrightarrow{#1} #3}

\renewcommand{\theenumii}{\theenumi.\arabic{enumii}.}

\begin{document}




\title{\system: A Message Broker System for Enabling Distributed, Internet-Scale IoT Applications over Heterogeneous Cloud Platforms}


\author{Sherif~Abdelwahab and Bechir Hamdaoui
%
~\\
\small Oregon State University, abdelwas,hamdaoui@eecs.orst.edu\\
\vspace{-20pt}
}

\maketitle
\pagenumbering{gobble}

\begin{abstract}
Excessive tail end-to-end latency occurs with conventional message brokers as a result of having massive numbers of geographically distributed devices communicate through a message broker. On the other hand, broker-less messaging systems, though ensure low latency, are highly dependent on the limitation of direct device-to-device (D2D) communication technologies, and cannot scale well as large numbers of resource-limited devices exchange messages.
In this paper, we propose~\system, a cloud-based message broker system that overcomes the limitations of conventional systems by enabling autonomous discovery, self-deployment, and online migration of message brokers across heterogeneous cloud platforms.
For each device, \system~provides a high capacity device cloning service that subscribes to device messages. The clones facilitate near-the-edge data analytics in resourceful cloud compute nodes.
Clones in \system~apply \protocol, an algorithm mimicking flocking-like behavior to allow clones to dynamically select and autonomously migrate to different heterogeneous cloud platforms in a distributed manner.

\end{abstract}

\begin{IEEEkeywords}
Message brokering architecture, cloud computing resource allocation, distributed IoT applications.
\end{IEEEkeywords}

\IEEEpeerreviewmaketitle

\section{Introduction}
\label{intro}
Many large-scale applications are sensitive to latency as they rely on messaging sub-systems between geographically distributed devices and cloud services. Even if $10\%$ of messages were delayed for longer than 150-300 ms, applications like remote-assisted surgery and real-time situation-awareness may not be feasible \cite{anvari2005impact, ramachandran2012large}. A bounded tail end-to-end latency is a cornerstone for the realization of large-scale \ac{iot} applications near the network edge \cite{rumble2011s, bonomi2014fog}.

When devices communicate through a middle message broker, successive packets queuing in multi-hop paths becomes a major source of latency. For example the average end-to-end latency of messages exchanged using a \redis~broker \cite{carlson2013redis} in a close amazon data-center is three times longer than deploying the same broker one-hop away from devices. Broker-less messaging using device-to-device communication does not necessarily solve the successive packets queuing problem. In \ac{iot} applications, a device communicates a large number of messages with many devices. Devices' limited processing and memory capacities become another major source of latency for large-scale distributed applications. Experiments show that direct device-to-device (D2D) messages can experience double the end-to-end latency when compared to brokering the messages through a one-hop away broker (see Section \ref{problem}).

When devices are cloned in a one-hop away cloudlet \cite{satyanarayanan2009case}, a device's clone can provide message brokering service so that interacting devices can communicate with low latency while allowing them to offload their computation to resourceful nodes. Of course, communicating through a one-hop away clone may still cause long \emph{tail end-to-end latency} - considering the 99-th percentile of computation plus communication latencies between clones/devices - when the broker service relays messages to distant devices. If a clone can measure: $1)$ messaging demand with other devices/clones, $2)$ the tail latency experienced by messages, and $3)$ the potential latency of other cloudlets/cloud platforms, clones can self-migrate between cloud platforms to always ensure a bounded weighted tail end-to-end latency. We show how autonomous clone migration can mimic birds flocking and prove that it is stable and achieves a tight latency that is $(1+\epsilon)-$far from optimal.

The use of cloudlets and dynamic service migration to solve latency problems are not new:
Cloudlets \cite{satyanarayanan2014cloudlets} reduce the single-hop latency from 0.5-1 seconds to tens of milliseconds, and technologies like MobiScud and FollowMe \cite{wang2015mobiscud,taleb2013follow} migrate clones to sustain an average single-hop \ac{rtt} at nearly 10 ms. Such schemes struggle to make optimal migration decisions despite using central control units as they: adopt too constraining migration metric (average single-hop latency) and trigger migration only if devices locations change \cite{urgaonkar2015dynamic}.
However, applications in fog computing \cite{bonomi2014fog} necessitate the deployment of inter-networking  clones in heterogeneous platforms (cloudlet/clouds) without centralized administration. In this fog environment, clones communicate with several geo-distributed devices and other clones where the \emph{tail weighted end-to-end latency} becomes the primary latency measure instead of the average \ac{rtt} of a single-hop.

We propose \system, a clone brokering system design that allows clones to self-discover and autonomously migrate to potential cloud hosting platforms according to self-measured weighted tail end-to-end latency, thereby stabilizing clone deployments and achieving low latency. \system~has four key features:

\begin{figure*}[ht]
\hfill
\subfigure[Experiments setup.]{\label{fig:motivexp}\includegraphics[width=0.36\textwidth]{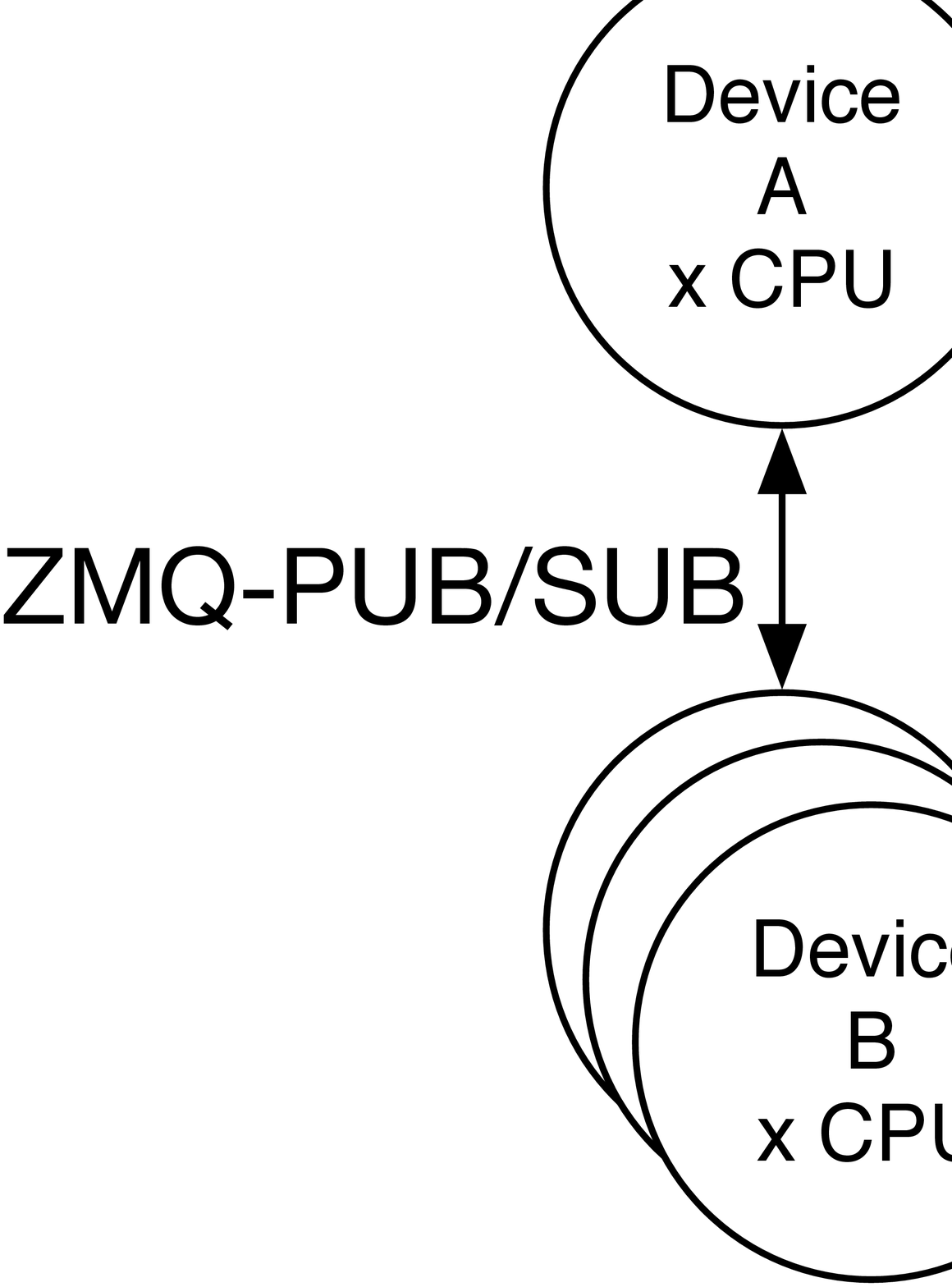}}
\hfill
\subfigure[CDF of latencies.]{\label{fig:edgelatency} \includegraphics[width=0.3\textwidth]{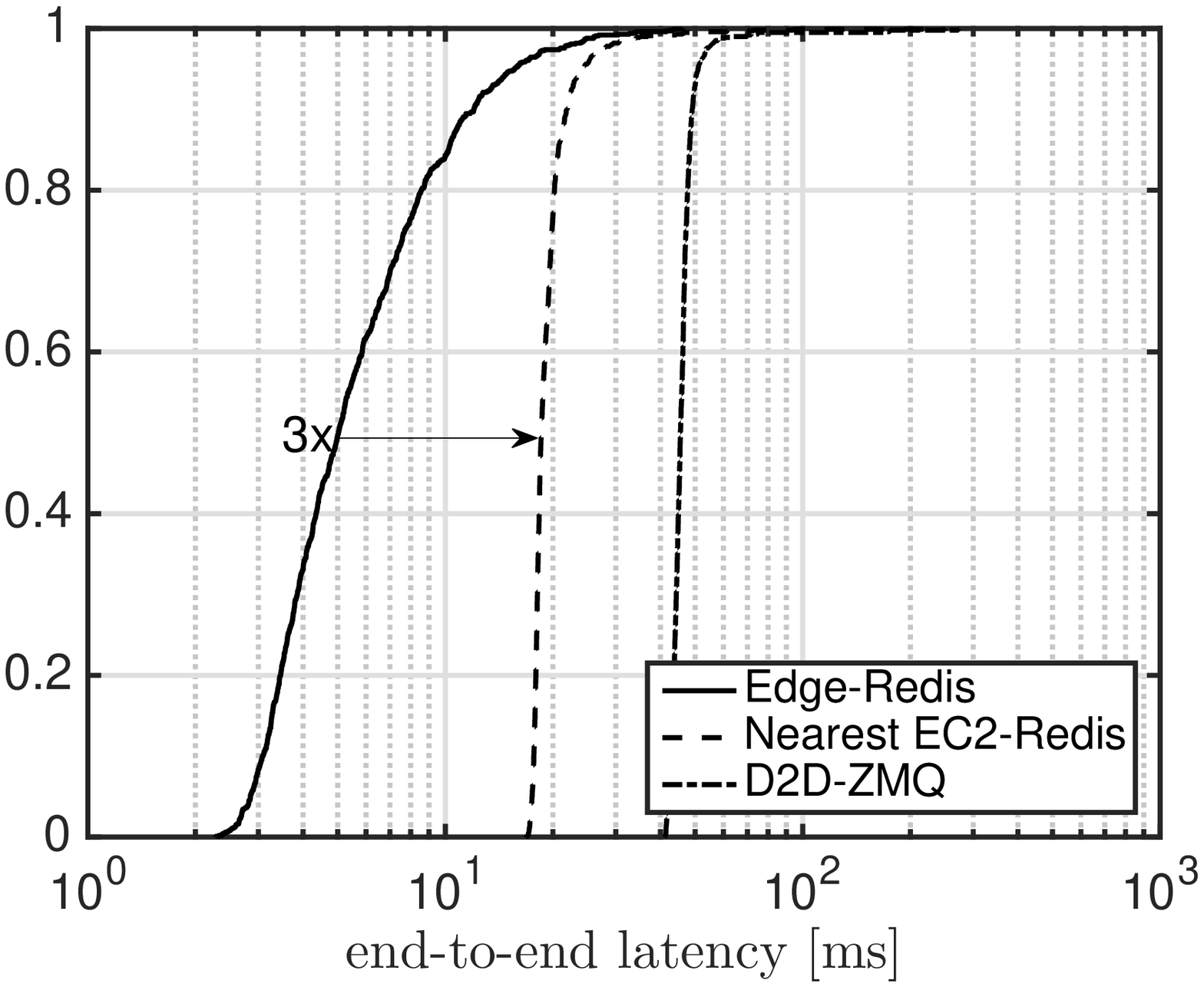}}
\hfill
\subfigure[Tail latency in D2D vs edge brokers.]{\label{fig:d2dlatency} \includegraphics[width=0.3\textwidth]{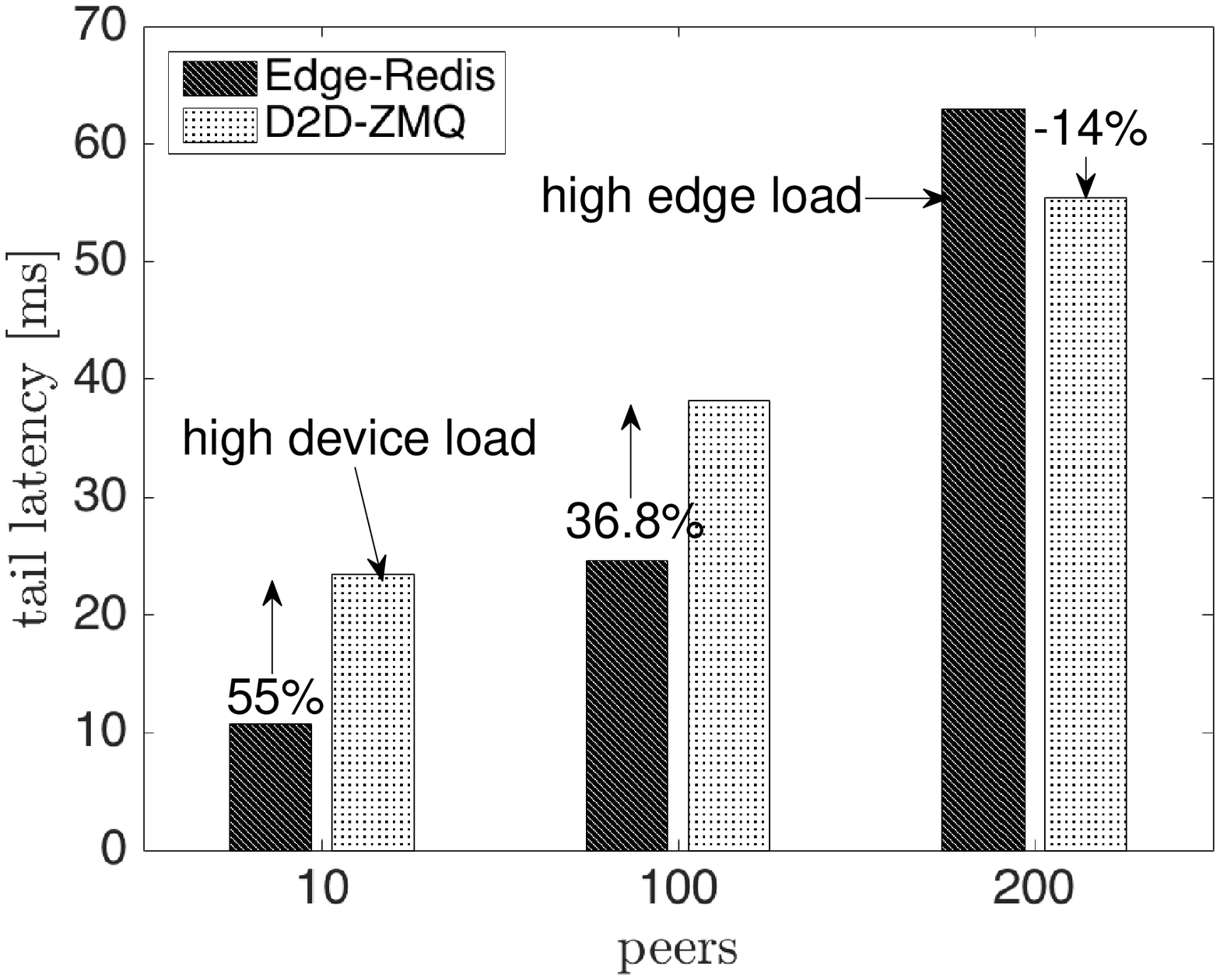}}
\hfill
\caption{Motivating experiments: the sources of latency in devices publish/subscribe communication.}
\end{figure*}

\begin{enumerate}
\item Reduces end-to-end delays that arise from multi-hop message queuing by deploying message brokers at the network edge, and while accounting for the devices' communication and relationship traffic patterns;


\item Ensures bounded message latency of IoT applications, outperforming conventional message brokers, like \redis~and \nats;
\item Autonomously discovers and migrates to heterogeneous, Internet-scale cloud/edge platforms without needing centralized monitoring and control;
\item Simple and requires no change to existing cloud platforms controllers.
\end{enumerate}

\section{Motivation and Challenges}
\label{problem}
We first show that multi-hop queuing along Internet paths is a major source of end-to-end latency for \ac{iot} applications. We then show that devices' limited compute and memory resources standstill against latency reduction by direct device-to-device communication.

\subsection{Sources of latency}

When a message broker is hosted in a multi-tenant cloud platform, the end-to-end latency degrades due to \emph{network interference}. Network interference occurs when the broker messages share: ingress/egress network I/O of its host, and one or more queues in the data-center network switches \cite{barker2010empirical}. Hosts and network resources become spontaneously congested by traffic-demanding applications. For a single-authority cloud, an operator can control network interference of latency-sensitive applications with: switching, routing, and queuing management policies, besides controlling contention for hosts' compute, memory, and I/O resources \cite{pu2013your,chiang2011tracon}.

Network interference is harder to control for \ac{iot} applications. As devices communicate using our cloud-hosted broker, messages share network resources of multi-hop paths with diverse and unmonitored traffic.  Multiple, unfederated authorities manage network resources along these paths, which makes it hard to enforce \emph{unified} traffic shaping or queuing management polices. Adding also variations in devices' traffic demand, communication pattern with other devices, and mobility, it becomes particularity hard to trace devices' traffic, delays, and infrastructure conditions to find optimal policies with centralized solutions. Multi-hop queuing along Internet paths can account for a 3x degradation in end-to-end latency on average.

\figurename~\ref{fig:motivexp} illustrates an experiment setup to quantify this latency degradation. We install a \redis~server in a \ac{vm}-instance in the nearest Amazon EC2 data-center (EC2-\redis), and install another \redis~server in a same capacity \ac{vm} in a host co-located with our WiFi access point (Edge-\redis). Our host runs other workloads. We emulate devices as simple processes running on another host that uses the same access point. We ensure that all \acp{vm} and hosts are time-synchronized with zero delay and jitter during the experiment execution time. A device emulator $A$ publishes $10K$ messages to either \redis~servers, and another device emulator $B$ subscribes to $A$'s messages. \figurename~\ref{fig:edgelatency} shows the \ac{cdf} of the end-to-end latency, measured as the time between receiving a message at $B$ and publishing it from $A$. The tail end-to-end latency for Edge-\redis~is $15.6 ms$, while it measured at $24.2 ms$ for EC2-\redis~accounting for 1.5x tail end-to-end latency improvement by avoiding multi-hop path to the closest EC2 instance and 3x improvement on average.

\subsection{Why broker-less is not always the answer?}

\begin{table}[h]
\centering
\begin{tabular}{ l | c | c }
  \bf{Benchmark} & $50\%$ & $99^{th}\%$ \\ \hline
  \redis, $1000$ messages & $523.2$ & $1,276.0 \mu s$ \\
  \zmq, $1000$ messages & $314.8$ & $647.6 \mu s$ \\
  \redis, $10,000$ messages & $620.1$ & $2,010.5 \mu s$ \\
  \zmq, $10,000$ messages & $320.3$ & $652.9 \mu s$ \\
  \hline
\end{tabular}
\caption{Median and $99^{th}$ end-to-end latency of \redis~and \zmq~measured under different loads (number of messages).}
\label{tab:rediszmq}
\end{table}

Direct device-to-device communication using broker-less message queues can thought of to be better than using message brokers. The obvious reasons for broker-less queues, such as \zmq~\cite{hintjens2013zeromq}, superiority are: their lightweight implementation, and their usage of minimal number of shared queues, switches, routers, and access points, between communicating devices. \tablename~\ref{tab:rediszmq} shows the median and tail end-to-end latency of \zmq~and \redis~under different loads, where \zmq~can deliver $10,000$ messages three times faster than \redis.

Unfortunately, if the devices are resource limited, the latency superiority of direct device-to-device communication is not always maintained. Let us return to our motivating experiment in \figurename~\ref{fig:motivexp}.
We limit the resources used by the device emulators using Linux \cgroups~such that a device emulator can use no more than $10\%$ of the CPU time compared to EC2-\redis~or Edge-\redis. \figurename~\ref{fig:edgelatency} shows that the average end-to-end latency of D2D-\zmq~is 7 times longer than Edge-\redis, and the tail end-to-end latency is 4 times longer. Several factors can contribute to this deteriorated performance including the wireless environment loading and implementation details of either \redis~or \zmq. However, the main factor that limits direct device-to-device latency is the limited compute resources of the devices emulators.

To emphasis this observation, we increase the number of the publishing device peers (i.e. number of subscribing device emulators) until the Edge-\redis~server becomes loaded. \figurename~\ref{fig:d2dlatency} shows the tail end-to-end latency for different numbers of peers. As we increase the number of peers, the latency superiority of the Edge-\redis~starts to diminish, until we reach the 200 peers points at which our host becomes loaded at $90\%$ utilization and the tail end-to-end latency of broker-less D2D-\zmq~becomes better by $14\%$. Broker-less device-to-device messaging is only better if a device computational resources are sufficiently large, which is an unrealistic assumption for most \ac{iot} devices.

\section{\system~System}
\label{design}
Our motivating experiments show that multi-hop queuing along Internet paths is a major source of tail end-to-end latency for cloud-based messaging systems, and that the latency improvement promise from device-to-device communication cannot  always be attained due to limited devices resources. \system~tackles multi-hop queuing by reducing the queuing of messages: primarily, if message brokers can self-deploy and migrate in cloudlets according to the communication pattern of the devices, then the impact of multi-hop queuing delay can be diminished. In the extreme case, if two resource-limited devices communicate through brokers in the same cloudlet using the same access point, we can achieve a finite minimal bound on the latency. Autonomous brokers migration is the foundation idea of \system.

In this section, we derive an intuitive design of \system~by which we bound weighted latency given an arbitrary network of heterogeneous cloud platforms. Although the stability and bounded performance of our design are intuitive, we solidify this intuition by relating the design to the theory of singleton weighted congestion games \cite{fotakis2002structure,fotakis2005selfish}, where we show that self-deploying clones reach a \ac{ne} and tightens the \ac{poa} of the weighted end-to-end delay.

\subsection{Network of clones}

To begin, we assume that devices communicate with each others according to \ac{iot} applications' requirements and form a social network of devices. Typically, the convergence of man-machine interactions in \ac{iot} will derive devices to form a social network \cite{atzori2011siot}. This network can form according to existing social network structure of users or according to the required communication among devices that is inherited from application design.

The idea of application design for \ac{iot} is simple. An application is modeled as a graph (e.g. \cite{hong2013mobile,ahmed2014graph, satish2014navigating} ). Each device participating in the execution of the \ac{iot} application \emph{publishes} its data to its brokering clone.
On the other hand, clones \emph{subscribe} to each other according to the application-modeled graph, which forms an overlay network of clones to enable the IoT application. Upon completion of their executions, clones may \emph{push} the results back to devices.

\begin{figure}
\centering
\includegraphics[width=0.49\textwidth]{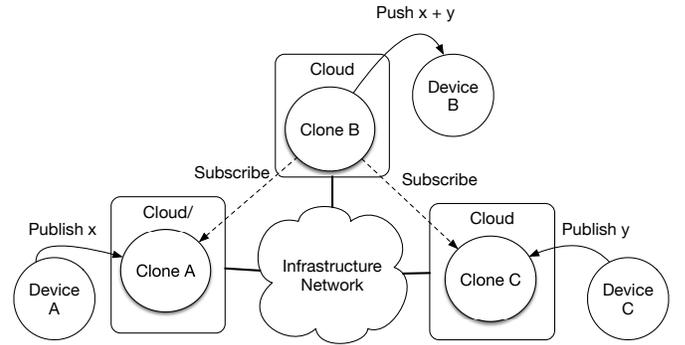}
\caption{Example overlay aggregation tree formed by device clones.}
\label{fig:overlay1}
\end{figure}

\figurename~\ref{fig:overlay1} illustrates a simple tree aggregation application for data retrieved from three devices. Device A and C publish their data $x$ and $y$ to their clones. As device $B$ is interested in the result $x+y$, clone B subscribes to data from clones A and C to retrieve $x$ and $y$, evaluates $x+y$, and pushes the result to its device. The advantages of using a pub/sub system for interacting between clones and the devices are: $1)$ providing an efficient messaging middleware to manage large-scale graph structures and multiple applications, $2)$ relying on the already in-place subscription and matching languages to effectively route information between devices and clones and inter-clones, and $3)$ simplifying the design of large-scale applications as overlay networks of and among the clones.

Generally speaking, the overlay network design of the clones is either structured or unstructured, and focuses mainly on minimizing a brokers fanout to minimize the communication between the clones. For example, topic-connected overlay networks are designed such that devices interested in the same topic are organized in a direct connected dissemination overlay \cite{chen2012generalized}. The overlay network forms the foundation for distributed pub/sub, and directly impacts the system scalability and application performance~\cite{baldoni2007efficient,chen2015algorithms, sun2013low}. We assume that an overlay topology of clones is given and we model it as a social network. We model the social network of clones as a graph $G=(V,P)$, where $V$ denotes the set of $n$ clones and $P$ denotes the set of all clone pairs such that $p=(i,j) \in P$ if the $i$-th and $j$-th clones communicate with each other.

\subsection{\system~architecture}

\begin{figure}
  \centering
\includegraphics[width=0.5\textwidth]{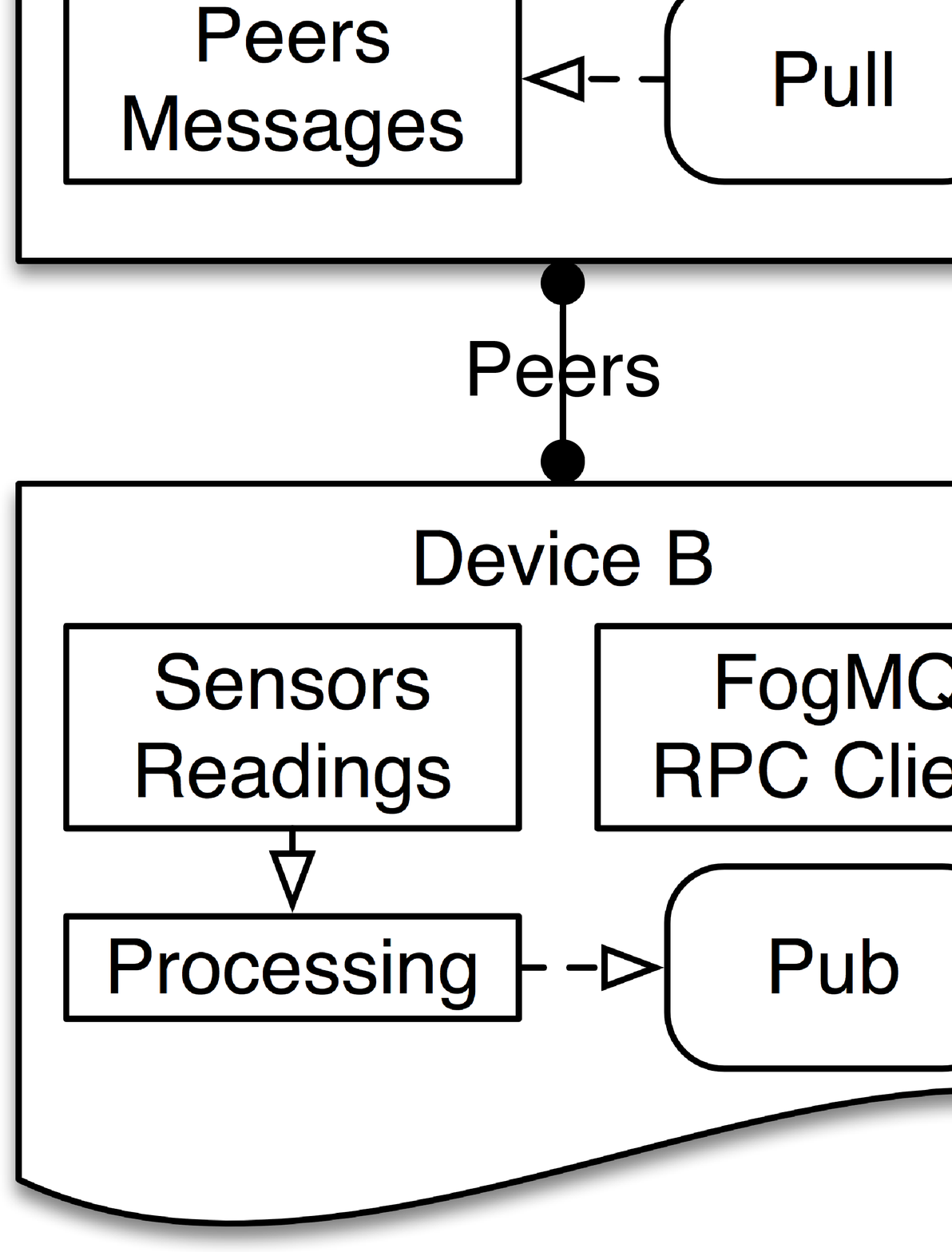}
  \caption{\system~Architecture}
  \label{fig:arch}
\end{figure}

We now describe how \system~initially creates device clones, as well as the \system's~architectural design tradeoffs. \figurename~\ref{fig:arch} illustrates the architectural elements of \system.

\system~initially clones a device at the closest cloud/cloudlet from a set $A$ of $m$ cloud/cloudlets that are available to all devices and that can communicate over the Internet. An RPC client in the device is responsible for the clone initiation and peer relationship definition with other devices by which a device participates in the execution of a distributed application described as an overlay network of clones. Each cloud/cloudlet runs \system~RPC server as a middleware around the cloud/cloudlet controller which implements different solutions that enable \system~to interact with heterogeneous cloud platforms (e.g. EC2, AWS, OpenStack cloudlet, or even a standalone host).  A typical approach that clients can use to initiate clones is to query a global geo-aware domain name service load balancer to retrieve the IP address of the nearest \system~RPC server. With the integration of cloud computing in cellular systems \cite{abdelwahab2015replisom}, devices can also use native cellular procedures to initiate clones in the device's nearest cellular site.

The \system~RPC server realizes the device clone in a cloud platform as a virtual machine, container, or native process, where processes are always a favorable design choice to avoid latency overhead. Recent \linpack benchmark \cite{dongarra2003linpack} shows that containers and native processes can achieve a comparable number of floating point arithmetic per second, at least $2.5$x greater than virtual machines. Despite that containers have better privacy and security advantage, as they provide a better administration, network, storage, and compute isolation, containers networking configuration can account for $30\mu s$ latency overhead when compared to that incurred by native processes \cite{felter2015updated}. As we will discuss later, implementing clones as processes has an advantage over both virtual machines and containers as they incur lesser migration overhead.

Once \system~creates a clone for a device, the clone subscribes to the devices' published messages that contain preprocessed sensors reading. Subscribing to the devices' messages eases clone migration processes as we will detail later. If a clone migrates from one cloud to the other, any changes to the clone IP address or network configuration become transparent to the device. Upon migration, the clone resubscribes to the devices' messages, allowing the device to continue publishing its messages without needing the clone to notify the device of its migration.

A clone can process its device's messages in its computation offloading module (see \figurename~\ref{fig:arch}) with high processing, memory, and storage capacities. The computation offloading module of the clone also executes distributed applications defined as overlay networks that interconnect several clones. To exchange messages between peer-clones of an overlay network, a clone creates a separate process for each of its peers. Each process subscribes to published messages from its corresponding peer-clone to make messages from peers available for the computation offloading module. If needed a clone pushes messages and/or computation results back to its device using a push/pull messaging pattern. \figurename~\ref{fig:arch} illustrate messages flow between different modules for a simple example in which Device A is a peer to Device B.

The overlay optimization module and the peer-to-peer routing module are responsible for optimizing the fan-out of overlay networks and the routing decisions as we described earlier. Although the design of these mechanisms are integral to the performance of the overall system, the details overlay design and routing optimization algorithms are orthogonal to the scope of this paper in which we focus on autonomous migration decision that minimize the tail end-to-end latency.

\subsection{Latency and peer-demand monitoring}

Each device clone self-monitors and characterizes the demands with its peers, and evaluates latencies with the assistance of the hosting cloud.
Let $d_{ij} \in \Rset^+$ denote the traffic demand between $i$ and $j$ and assume that $d_{ij} = d_{ji}$.
Let $x_i \in A$ denote the cloud that hosts $i$ and $l(x_i, x_j) > 0$ be the average latency between $i$ and $j$ when hosted at $x_i$ and $x_j$, respectively (Note: if $i$ and $j$ are hosted at the same cloud, $x_i = x_j$).  We assume that $l$ is reciprocal and monotonic. Therefore, $l(x_i, x_j) = l(x_j, x_i)$ and there is an entirely nondecreasing order of $A \rightarrow A'$ such that for any consecutive $x_i, x_i' \in A'$, $l(x_i, x_j) \leq l(x_i', x_j)$. The reciprocity condition ensures that measured latencies are aligned with peer-\acp{vm} and imitates the alignment rule in bird flocking. We model $l(x_i, x_j) = \tau (x_i, x_j) + \rho (x_i) + \rho(x_j)$, where $\tau (x_i, x_j)$ is the average packet latency between $x_i$ and $x_j$, and $\tau (x_i, x_j) = \tau (x_j, x_i)$. The quantity $\rho(x)$ is the average processing delay of $x$ modeled as: $\rho (x) = \delta \sum_{i \in V: x_i = x} \sum_{j\in V} d_{ij} / (\gamma(x) - \sum_{i \in V: x_i = x} \sum_{j\in V} d_{ij}),$ where $\delta$ is an arbitrary delay constant and $\gamma(x)$ denotes the capacity of $x$ to handle all demanded traffic of its hosted \acp{vm}.

\subsection{Learning new targets}

Each cloud runs \system~RPC server as a middleware. The \system~servers in different clouds form a peer-to-peer network that evolves autonomously. Bootstrap nodes assist newly joined \system~servers to discover other servers. Gossip protocol is used to spread information about new \system~servers.

\section{\protocol---An Adaptive Clone Migration
Algorithm}
\label{flock}
Clones should adapt themselves to changes in the infrastructure network interconnecting the heterogenous cloud platforms, and should change according to the network state, structure, and applications' requirements. We propose an adaptive, fully distributed algorithm for dynamic cloud selection. The algorithm allows each \ac{vm} $i$ to learn a set $A_i \subseteq A$ (referred to as $i$'s strategy set) from its hosting cloud $x_i$, and to autonomously select its hosting cloud based on local measurements only.
Every cloud $x$ updates its weight $w_x = \sum_{i: x_i=x} u_i(x)$ and broadcasts a monotonic, non-negative regularization function $f(w_x): \Rset^+ \rightarrow \Rset^+$ with $\alpha < f(w_x) < 1$ for $\alpha > 0$ to each \ac{vm} hosted at $x$.  As \acp{vm} can each only be hosted by one cloud and all have access to the same set of strategies, we model the clone migration problem as a singleton symmetric weighted congestion game that minimizes the social cost $C(\sigma) = \sum_{x \in A} w_x f(w_x)$. If $f(w_x) \approx 1 $, this game model approximates to minimizing $\sum_{i \in V} u_i(x_i)$. Let $\migrate{i}{x}{y}$ denote that a \ac{vm} $i$ migrates from cloud $x$ to cloud $y$ and let $\eta \leq 1$ denote a design threshold. We now describe our proposed clone migration algorithm:
\vspace{-5pt}
\begin{algorithm}[H]
\centering
\begin{algorithmic}[1]
  \renewcommand{\algorithmicrequire}{\textit{Initialization:}}
  \renewcommand{\algorithmicensure}{\textit{Ensure:}}
  \REQUIRE Each clone $i \in V$ runs at a cloud $x \in A$.
  \ENSURE  A Nash equilibrium outcome $\sigma$.
  \STATE During round $t$, do in parallel: for all $i \in V$
  \STATE $i$ solicits its current set $A_i$ from $x$.
  \STATE $i$ randomly selects a target cloud $y \in A_i$.
  \IF { $u_i(y) f(w_y + u_i(y)) \leq  \eta  u_i(x) f(w_x - u_i(x))$}
    \STATE $\migrate{i}{x}{y}$
  \ENDIF
\end{algorithmic}
\caption*{\protocol: Autonomous \ac{vm} migration protocol.}
\end{algorithm}

The following theoretical results have been proven in our work~\cite{flocking-arxiv}, and are included here for completeness.

\begin{theorem}
\label{theorem1} (Theorem 3.1~in~\cite{flocking-arxiv})
  \protocol~converges to a Nash equilibrium outcome.
\end{theorem}

\begin{lemma}
\label{lemma1} (Lemma 3.2 in~\cite{flocking-arxiv})
The social value of \protocol~has a perfect \ac{poa} at most $\lambda/ (1-\varepsilon)$ if for $\varepsilon < 1$ and $\lambda > 1 - \varepsilon$ the regularization function satisfies
$w^* f(w + w^*) \leq \lambda  w^* f(w^*) + \varepsilon w f(w) $, where $w \geq 0$ and $w^* > 0$.
\end{lemma}

\begin{theorem}
\label{theorem2} (Theorem 3.3 in~\cite{flocking-arxiv})
The regularization function $f(w) = exp(-1/(w+a))$ tightens the \ac{poa} to $1+\epsilon$ for a sufficiently large constant $a$ and reduces the game to the original \ac{vm} migration problem, i.e. minimizing $\sum_{i} u_i(x_i)$.
\end{theorem}

We now provide some simulation results, borrowed from~\cite{flocking-arxiv} for completeness, to have an initial sense of how well \protocol~performs in terms of convergence and achievable \ac{poa}; more results on the performance of \protocol~can be found in~\cite{flocking-arxiv}. Clouds are modelled as a complete graph with inter-cloud latency $\tau \sim \text{Uniform}(10, 100)$ and cloud capacity $\gamma \sim \text{Uniform}(50, 100)$. We model peer-to-peer clone relations as a binomial graph with $d \sim \text{Uniform}(1, 10)$.

\begin{figure}[htp]
  \centering
  \includegraphics[width=0.45\textwidth]{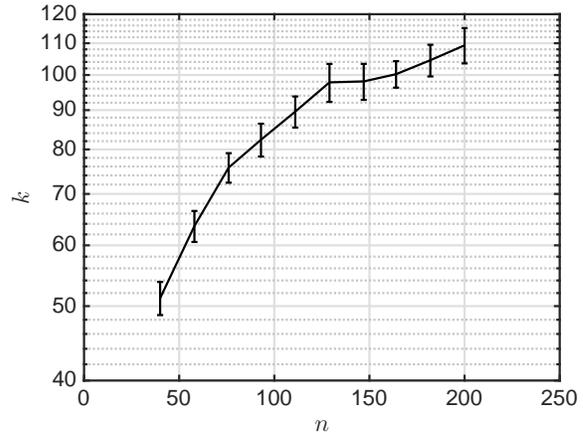}
  \caption{\protocol~convergence: $a=9$, $\eta=0.9$, and $\#$ of clouds $=37$.
  }
  \label{convergence}
\end{figure}

\figurename~\ref{convergence} shows the average number of rounds, $k$, needed for the algorithm to converge to a Nash equilibrium at $95\%$-confidence interval with $0.1$ error.
Observe that although the worst case of $k$ is $O(n \log (n f_{max}))$ where $f_{max}$ is the maximum value of the regularization function $f$ \cite{chien2007convergence}, the figure shows that \protocol~scales better than $O(n)$ on average.

\begin{figure}[htp]
  \centering
  \includegraphics[width=0.45\textwidth]{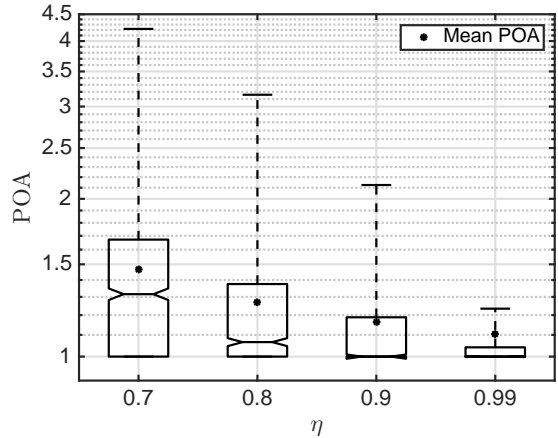}
  \caption{$a=9$, $n=8$, and $\#$ of clouds $=5$.
  %
  }
  \label{fig:poa}
\end{figure}

\figurename~\ref{fig:poa} shows the \ac{poa}~under different values of $\eta$. The optimal solution is implemented as brute-force in simulations to evaluate the minimum social value among all possible device clones-to-cloud assignments. Observe that as $\eta$ approaches $1$, the maximum achieved \ac{poa} matches the theoretical value of $1.21$. In practical systems, $\eta$ should not be too close to $1$, so as to avoid migration oscillations, where clones keep migrating without much improvement.
%

\section{Related Work}
\label{background}
We propose autonomous brokering clones for designing large-scale distributed pub/sub systems as a major mechanism that complements existing techniques to minimize the tail end-to-end messaging latency. Researches focus on three main techniques for the development of simple, scalable, and resource economic pub/sub systems: $1)$ content-centric above layer-3 routing between brokers (e.g. \cite{canas2015graps,carvalho2005scalable,an2015wide,chen2015design,rao2015towards}), $2)$ overlay brokers network topologies designs (e.g. \cite{siegemund2015self,chen2015weighted, teranishi2015scalable,chen2015algorithms,onus2011minimum}), and $3)$ content-centric in-network caching (e.g.\cite{diallo2011leveraging, cho2012wave, matos2010stan, matos2010stan}).

Distributed pub/sub systems organize brokers, devices, or routing functions as an overlays and sub-overlays at the application layer. Upon constructing an efficient overlay network, routing protocols above layer-3 build minimum-cost message dissemination paths to deliver messages to subscribers according to specific topic-interest. Caching policies replicate clones' contents closer to devices interested in a content for faster repetitive publishing. For a given routing, overlay topology, and caching mechanisms, \system~ensures that these mechanisms achieve their full potentials by self-reorganizing the deployment of brokers through migrations in heterogeneous, unmanaged, and dynamic cloud environments. Unlike widely adopted centralized systems (e.g. \redis), \system~suits the large-scale applications and use cases of \ac{iot} and avoids the limitations of broker-less systems (e.g. \zmq).

\comment{
\subsection{Overlay deployment}

Organizing broker overlay topologies is NP hard \cite{jaeger2007self}, and is related to the virtual network embedding problem in large-scale dynamic networks were existing embedding solutions become inapplicable. Given a broker network overlay, an efficient deployment embeds this overlay network onto an existing infrastructure network to minimize the tail end-to-end latency for message dissemination as our main goal.

Existing embedding solution are limited in their applicability in overlay deployment

Adapts the structure of the broker network by allowing a broker to exchange its caches information with other brokers that have common interest in the content of the first using consensus protocols. Unlike this content centric model, we rely on the advancement in process and virtual machine migrations to actually adapt the topology of the brokers and prove using game theory that our algorithm is very close to optimal despite the hardness of the problem.

Unlike content-based routing and overlay network design in pub/sub system \cite{majumder2009scalable}, our proposed scheme is oblivious from either layer-3 or layer-7 routing protocol decision taken by the pub/sub subsystems. Whatever the performance of the routing mechanism is used our broker still self-migrate to minimize their latency according to actual network performance. Improvement in routing and overlay design between brokers will be benefitiional to our scheme but it is an orthogonal domain of improvement. For example, SDN based approaches can integrate content-centric routing decision in SDN controllers \cite{tariq2014pleroma}.

\subsection{Migration solutions}
}

Existing migration solutions are limited in their applicability to minimize the weighted end-to-end latency.
Several existing solutions rely on a system-wide central controller to manage the states of clones, devices, and physical resources of cloud platforms \cite{urgaonkar2015dynamic, eramo2014study}.
For the considered fog environment, these solutions lack scalability for an Internet-sized network without relaxations that potentially compromise solutions quality.

Consider \ac{mdp} based solutions.
\ac{mdp} requires a central server to collect statistics of devices mobility, clones demands, and clouds connectivity and utilization.
This server also executes the value iteration algorithm to evaluate an optimal  migration policy \cite{urgaonkar2015dynamic, eramo2014study, chen2014distributed}.
It is intractable to model all possible states of clones and their hosting platforms; hence it is common to discretize states measurements to relax the complexity of the policy optimization algorithms \cite{chen2014distributed, urgaonkar2015dynamic}.
This compromises the solutions quality.

Game-theoretic approaches potentially decentralize the migration algorithms and improve their scalability.
However existing game-theoretic solutions provide an unbounded \ac{poa}~\cite{xiao2015solution}.
We cannot use them - as they are - and guarantee optimal or close to optimal weighted end-to-end latency.
Finally, existing migration solutions serve specialized cloud providers' objectives (e.g. energy, load, and cost) to profitably manage providers' infrastructures \cite{duong2014joint, xiao2015solution}.
\emph{The existing models do not capture network latency between clones that are executing distributed \ac{iot} applications.} Unlike existing solutions, \system~adopts a simple autonomous migration protocol that is stable and bounds the tail end-to-end latency $(1-\epsilon)$ far from optimal.

\section{Conclusion and Future Work}
\label{conclusion}
We proposed \system, a cloud-based, message broker system, composed of an architecture and an online migration algorithm, that enables autonomous discovery, self-deployment, and online migration of message brokers across heterogeneous cloud platforms. The migration algorithm, called \protocol, enables autonomous discovery of and migration to heterogeneous cloud/edge platforms in a (i) decentralized manner and (ii) without requiring changes to existing cloud platform controllers. The proposed architecture enables the deployment of message brokers (i) at the edge clouds (i.e., cloudlets) near the end-user devices, and (ii) while accounting for the devices' communication and relationship traffic patterns.

An implementation of \system~on real cloud platforms is currently underway. In this implementation, clones are implemented as processes, Redis key-value store as a device registry, and edge clouds as Linux VMs. Our implementation-based performance evaluation of our proposed \system~system, a clone-based architecture design and an online clone migration algorithm, will be published when available.

\section*{Acknowledgement}
This work was made possible by NPRP grant \# NPRP 5- 319-2-121 from the Qatar National Research Fund (a member of Qatar Foundation). The statements made herein are solely the responsibility of the authors.


\bibliographystyle{IEEEtran}
\bibliography{IEEEabrv,./references}

\end{document}